# Robust Portfolio Design and Stock Price Prediction Using an Optimized LSTM Model

Jaydip Sen
*Dept of Data Science*
*Praxis Business School*
Kolkata, INDIA
email: jaydip.sen@acm.org

Saikat Mondal
*Dept of Data Science*
*Praxis Business School*
Kolkata, INDIA
email: saikatmondal15@gmail.com

Gourab Nath
*Dept of Data Science*
*Praxis Business School*
Bangalore, INDIA
email: gourab@praxis.ac.in

*Abstract*— Accurate prediction of future prices of stocks is a difficult task to perform. Even more challenging is to design an optimized portfolio with weights allocated to the stocks in a way that optimizes its return and the risk. This paper presents a systematic approach towards building two types of portfolios, optimum risk, and eigen, for four critical economic sectors of India. The prices of the stocks are extracted from the web from Jan 1, 2016, to Dec 31, 2020. Sector-wise portfolios are built based on their ten most significant stocks. An LSTM model is also designed for predicting future stock prices. Six months after the construction of the portfolios, i.e., on Jul 1, 2021, the actual returns and the LSTM-predicted returns for the portfolios are computed. A comparison of the predicted and the actual returns indicate a high accuracy level of the LSTM model.

## I. Introduction

The task of designing optimum and robust portfolios has always been considered a research problem of intense interest among quantitative and statistical financial analysts and researchers. A portfolio is said to be optimal when it allocates weights to a set of stocks in such a way that the return and risk associated with the portfolio are traded-off in the best possible manner. Markowitz, in his seminal work, proposed an approach called "the mean-variance optimization approach", which is based on the mean and covariance matrix of the returns [1]. While identifying an optimal "mean-variance" portfolio belongs to the NP-hard class, following Markowitz's work, several propositions have been made by researchers for portfolio optimization and stock price prediction. Among the proposed methods in the literature for stock price prediction, multivariate regression, ARIMA, VAR, time series forecasting, and learning-based approaches are quite popular. However, since it is extremely difficult to accurately estimate the future prices of stocks, estimation of the expected returns of a stock from its historical prices is invariably error-prone. Hence, it is a popular practice to use either a minimum variance portfolio or an optimum risk portfolio with the maximum Sharpe Ratio as better proxies for the expected returns.

This paper presents a step-by-step approach towards designing robust and efficient portfolios by choosing stocks from four critical sectors of the National Stock Exchange (NSE) of India. Based on the report of NSE on Jul 30, 2021, the ten most significant stocks of each of the four sectors are first identified [2]. Based on the historical prices of the forty stocks from Jan 1, 2016, to Dec 31, 2020, efficient portfolios are designed for the sectors optimizing their risks and returns and exploiting their principal components. To augment the process of portfolio construction, an LSTM model is designed for predicting future stock prices and future returns of the portfolios. The actual returns of the portfolios and the predicted returns by the LSTM model are compared six months after the construction of the portfolios to evaluate the accuracy of the predictive model. Further, the actual returns and volatilities reflect the current return and the risk associated with each sector studied in this work.

The contribution of this work is threefold. First, the work proposes two different approaches towards portfolio construction, the eigen portfolios, and the optimum risk portfolios. These two methods of portfolio design are applied to stocks of four critical sectors listed in the NSE. These portfolios will surely be a good guide for the investors in making effective and profitable investment decisions. Second, it proposes an efficient and optimized design of an LSTM architecture for predicting future prices of stocks for designing robust portfolios. Finally, the actual returns of the portfolios indicate the current profitability and volatility of the four sectors.

The paper is organized as follows. In Section II, some of the existing works on portfolio management and stock price prediction are discussed briefly. Section III provides the details of the data used and the methodology followed. Section IV discusses the design of the LSTM regression model. Section V discusses the results of different portfolios and the predictions of the future stock prices made by the LSTM model. Section VI concludes the paper.

## II. Related Work

Due to the challenging nature of the problem related to stock price prediction and robust portfolio design, and since these applications find impactful use cases in the real world, several propositions exist in the literature of these research areas. The use of predictive models built on learning algorithms and deep neural architectures for stock price prediction has been quite popular [3-6]. Hybrid models are proposed integrating learning-based algorithms and architectures with the sentiments in the unstructured data on the social web [7-9]. Several adaptations of Markowitz's minimum variance approach are proposed by researchers including purchase limit constraints and cardinality constraints. *Generalized autoregressive conditional heteroscedasticity* (GARCH) is a common approach for estimating the future volatilities of stocks and portfolios [10]. The use of metaheuristics in solving multi-objective optimization problems for portfolio design, eigen portfolios using principal component analysis, and linear and non-linear programming-based approaches are proposed by some researchers [11-13]. Further, *fuzzy logic*, *genetic algorithms* (GAs), *particle swarm optimization* (PSO) are also some of the popular approaches for portfolio design [14-15].

The current work presents two intrinsically distinct methods to portfolio design, (i) the optimum risk portfolio and (ii) the eigen portfolio, for robust designing portfolios for four important economic sectors of India. Based on the stock prices from Jan 1, 2016, to Dec 31, 2020, eight portfolios are designed, two for each sector. An LSTM model is then built for predicting the future prices of the stocks of each portfolio. Six months after the portfolio construction, the actual return for each portfolio and the return predicted by the LSTM model are computed to analyze the profitability of each sector and the prediction accuracy of the LSTM model.

## III. Data and Methodology

In Section I, it was pointed out that the primary objective of the work is to build robust portfolios for four critical sectors of the Indian economy. The second goal is to evaluate the prediction accuracy of the proposed LSTM model in predicting the future stock prices and future returns and risks associated with each portfolio. The return-risk analysis also provides one with insights into the current profitability and risk involved in each of the sectors analyzed. The Python programming language has been used in designing the portfolios and the LSTM model. The Tensorflow and Keras frameworks are also used. In the following, the seven-step approach followed in the design process is discussed.

### A. Choosing the Sectors

Four important sectors are chosen from the NSE, India. The chosen sectors are the following: *financial services*, *oil and gas*, *pharma*, and *public sector unit* (PSU) banks. The ten most significant stocks are identified for each sector, based on their contributions to the derivation of the sectoral index. The significant stocks are identified based on the report published by NSE on Jun 30, 2021 [2].

### B. Data Acquisition

For each sector, the prices of the top stocks are extracted using the *DataReader* function of the *data* sub-module of the *pandas_datareader* module in Python. The prices are extracted from the Yahoo Finance site, from Jan 1, 2016, to Jun 1, 2021. The stock price data from Jan 1, 2016, to Dec 31, 2020, are used for building the portfolios, while the portfolios are tested for their return on Jun 1, 2021. The current work is a univariate analysis, and hence, the variable *close* is chosen as the variable of interest, ignoring the remaining variables.

### C. Deriving the Return and Volatility of the Stocks

The daily returns are computed for each stock as the changes in the successive close values in percentage. The Python function pct_change is used for computing the daily returns. The variance, and hence the standard deviations of the daily return values are then derived to arrive at the daily volatility values for each stock. Assuming that there are 250 operational days in a calendar year, the annual volatilities are for the stock are computed by multiplying the daily volatilities by a factor of the square root of 250. The risk involved in stock is manifested in its annual volatility figure.

### D. Designing the Minimum Risk Portfolios

After computing the annual returns and risks (i.e., volatilities) of all the stocks, the minimum risk portfolios are designed for the four sectors. The minimum risk portfolio is the portfolio that has minimum variance associated with its constituent stocks. To identify the minimum variance portfolio for a sector, the efficient frontier of a large number of candidate portfolios is first plotted. The efficient frontier depicts the contour or locus containing portfolio points that yield a maximum return for a given value of risk, or they involve the minimum risk for a given value of the return. For an efficient frontier, the return and the risk are plotted along the y-axis and the x-axis, respectively. It is evident that the left-most point on a given efficient frontier depicts the minimum risk portfolio. In the current work, the efficient frontier for each sector is identified by randomly assigning weights to its constituent stocks and iterating such random assignment 10000 times over a loop so that 10000 such portfolios are designed. These portfolios are plotted on a two-dimensional space, and the left-most point along the x-axis is identified to determine the minimum risk (or minimum variance) portfolio. The return and risk for a portfolio are derived Using (1) and (2). In (1), *Ret* depicts the return of a portfolio with n stocks $S_1$, $S_2$, ...$S_n$, with respective weights $w_i$'s.

$$Ret = w_1 Ret(S_1) + w_2 Ret(S_2) + \cdots + w_n Ret(S_n) \quad (1)$$

In (1), *Ret* represents the return of a portfolio consisting of *n* stocks, which are represented as $S_1$, $S_2$, ...$S_n$, while $w_i$'s are their corresponding weights. The variance of a portfolio is given by (2).

$$V = \sum_{i=1}^{n} w_i s_i^2 + 2 * \sum_{i,j} w_i * w_j * covar(i,j) \quad (2)$$

In (2), *V*, $w_i$, and $s_i$ represent the variance of a portfolio, the weight of the $i^{th}$ stock, and its standard deviation. The covariance between the prices of the $i^{th}$ and the $j^{th}$ stock is represented as *covar(i, j)*.

### E. Identifying the Minimum Risk Portfolios

Since the minimum risk portfolios are rarely adopted by the investors in the real-word due to the low returns that they usually yield, a trade-off between the risk and return is carried out to optimize the return and risk. This trade-off leads to the design of optimum risk portfolios. For optimizing the risk, a metric called Sharpe Ratio (SR) is used. SR of a portfolio is the ratio of the difference between its return and that of a risk-free portfolio, to its standard deviation. as Sharpe Ratio is used, which is given by (3).

$$SR = \frac{Ret_{current} - Ret_{free}}{STD_{current}} \quad (3)$$

In (3), $Ret_{curent}$, $R_{free}$, and $STD_{current}$ represent the returns of the current and the risk-free portfolios, and the risk of the current portfolio, respectively. A risk-free portfolio is assumed to have a volatility of one percent. For a given set of stocks, the *optimum-risk portfolio* is designed to maximize the *Sharpe Ratio*. Once, the optimum risk portfolio is determined the corresponding weights assigned to the individual stocks are available. The Python function *idxmax* is used to identify the portfolio with the maximum SR value.

### F. Building the Eigen Portfolios

Designing eigen portfolios involves the concept of *principal component analysis* (PCA), a well-known dimensionality reduction method based on unsupervised learning. PCA retains the intrinsic variance in the data while reducing the number of dimensions. The principal components in the training data of the stock prices are determined using the PCA function defined in the *sklearn* library of Python. To retain 80% of the variance in the original stock price data, it is found that a minimum of five components is needed from the ten stocks. The components generated by the PCA function are orthogonal to each other, and their power of explanation of the variance in the data decreases with a higher component number. In other words, the first component explains the maximum percentage of the total variance. The component loading of the five principal components on each of the ten stocks reflects the weights allocated to the stocks in building the candidate eigen portfolios. Finally, the portfolio yielding the maximum Sharpe Ratio among the candidates is selected as the best eigen portfolio. A Python function is used iterating over a loop for deriving the weights assigned to the five principal components and in identifying the best candidate eigen portfolio [12].

*G. Predicted and Actual Returns and Risks of Portfolios*

Using the training dataset of the stock price from Jan 1, 2016, to Dec 31, 2020, two portfolios are designed for each sector, an optimal risk portfolio, and an eigen portfolio. On Jan 1, 2021, a fictitious investor is created who invests an amount of Indian Rupees (INR) of 100,000, for each sector based on the recommendation of the optimal risk portfolio structure for the corresponding sector. Note that the amount of INR 100,000 is just for illustrative purposes only. Our analysis will not be affected either by the currency or by the amount. To compute the future values of the stock prices and hence to predict the future value of the portfolio, a regression model is designed based on LSTM deep learning architecture. On May 31, 2021, using the LSTM model, the stock prices for June 1, 2021, are predicted (i.e., a forecast horizon of one day is used). Based on the predicted stock values, the predicted rate of return for each portfolio is determined. And finally, on June 1, 2021, when the actual prices of the stocks are known, the actual rates of return are derived. The predicted and actual rates of return for the portfolios are compared to evaluate the profitability of the portfolios and the accuracy of the LSTM model.

IV. THE LSTM MODEL

As explained in Section III, the stock prices are predicted with a forecast horizon of one day, using an LSTM deep learning model. This section presents the details of the architecture and the choice of various parameters in the model design. A very brief discussion on the fundamentals of LSTM networks and the effectiveness of these networks in interpreting sequential data is first discussed before the details of the model design are presented.

LSTM is an extended and advanced, *recurrent neural network* (RNN) with a high capability of interpreting and predicting future values of sequential data like time series of stock prices or text [16]. LSTM networks are able to maintain their state information in some specially designed memory cells or gates. The networks carry out an aggregation operation on the historical state stored in the *forget gates* with the current state information to compute the future state information. The information available at the current time slot is received at the *input gates*. Using the results of aggregation at the *forget gates* and the *input gates*, the network predicts the target variable's value for the next round. The predicted value is available at the *output gates* [16].

For predicting the stock prices for the next day, and LSTM model is designed and fine-tuned. The design of the model is exhibited in Fig. 1. The model uses daily *close* prices of the stock of the past 50 days as the input. The input data of 50 days with a single feature (i.e., *close* values) is represented by the data shape of (50, 1). The input layer forwards the data to the first LSTM layer. The LSTM layer is composed of 256 nodes. The output from the LSTM layer has a shape of (50, 256). Thus, each node of the LSTM layer extracts 256 features from every record in the input data. A dropout layer is used after the first LSTM layer that randomly switches off the output of thirty percent of the nodes in the LSTM to avoid model overfitting. Another LSTM layer with the same architecture as the previous one receives the output from the first and applies a dropout rate of thirty percent. A dense layer with 256 nodes receives the results from the second LSTM layer. The output of the dense layer produces the predicted *close* price. The forecast horizon may be adjusted to different values by changing a tunable parameter. A forecast horizon of one day is used so that a prediction for the following is made. The model is trained using a batch size of 64, and 100 epochs are used. While the sigmoid function is used for activation at the final output layer, the ReLU activation is deployed at all other layers. The loss and the accuracy during training and validation are measured using the *Huber loss function* and the *mean absolute error* function, respectively. The hyperparameter values used in the network design are all chosen based on the grid search method.

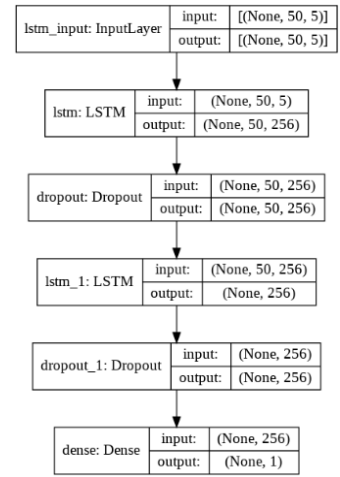

Fig. 1. The schematic diagram of the LSTM model

V. EXPERIMENTAL RESULTS

This section presents the detailed results of the performances of the portfolios and their analysis. The four sectors of the Indian stock market we choose are (i) *financial services*, (ii) *oil and gas*, (iii) *pharma*, and (iv) *public sector unit (PSU) banks,* and (vii) *realty*. The portfolios and the LSTM models are implemented using Python, and its associated libraries in TensorFlow and Keras. The models are trained and validated on the GPU environment of Google for faster execution and processing. The execution time of each epoch was three seconds approximately.

*A. Financial Services Sector*

The ten critical stocks of the financial services sector with their corresponding weights used for deriving the overall index of the sector based on the NSE report published on Jun 30, 2021, are HDFC Bank (HDB): 24.41, Housing Development Finance Corporation (HDF): 16.167, ICICI Bank Motors (ICB): 16.31, and Kotak Mahindra Bank (KTB): 9.35, Axis Bank (AXB): 7.19, State Bank of India (SBI): 6.01,

Bajaj Finance (BJF): 5.97, Baja Finserv (BFS): 2.73, HDFC Life Insurance (HLI): 2.12, and SBI Life Insurance (SLI): 1.66 [2]. We assume an imaginary investor who invested total capital of INR 100000 on Jan 1, 2021. The six months' returns are computed on Jul 1, 2021, for the optimum risk and the eigen portfolios. The LSTM model is used for predicting the stock prices of Jul 1, and the predicted return is compared with the actual return of the optimum risk portfolio.

TABLE I. ACTUAL RETURN OF OPT RISK PORTFOLIO (FIN SERV SEC)

| Stock | Wts | Date: Jan 1, 2021 | | | Date: Jul 1, 2021 | |
|---|---|---|---|---|---|---|
| | | Amnt Invstd | Act Price | No of Stocks | Act Price | Act Value |
| HDB | 0.2297 | 22970 | 1425 | 16.12 | 1487 | 23969 |
| HDF | 0.0149 | 1490 | 2569 | 0.58 | 2459 | 1426 |
| ICB | 0.0914 | 9140 | 528 | 17.31 | 631 | 10923 |
| KTB | 0.0416 | 4160 | 1994 | 2.09 | 1716 | 3580 |
| AXB | 0.0676 | 6760 | 624 | 10.83 | 746 | 8082 |
| SBI | 0.0086 | 860 | 279 | 3.08 | 420 | 1295 |
| BJF | 0.4010 | 40100 | 5280 | 7.59 | 5967 | 45318 |
| BFS | 0.0149 | 1490 | 8870 | 0.17 | 11816 | 1985 |
| HLI | 0.0075 | 760 | 678 | 1.11 | 686 | 761 |
| SLI | 0.1227 | 12270 | 895 | 13.71 | 1007 | 13806 |
| Total | | 100000 | | | | 111145 |
| Actual Return: 11.15 % | | | | | | |

TABLE II. ACTUAL RETURN OF EIGEN PORTFOLIO (FIN SERV SEC)

| Stock | Wts | Date: Jan 1, 2021 | | | Date: Jul 1, 2021 | |
|---|---|---|---|---|---|---|
| | | Amnt Invstd | Act Price | No of Stocks | Act Price | Act Value |
| HDB | 0.1100 | 11000 | 1425 | 7.72 | 1487 | 11479 |
| HDF | 0.1100 | 11000 | 2569 | 4.28 | 2459 | 10529 |
| ICB | 0.1100 | 11000 | 528 | 20.83 | 631 | 13146 |
| KTB | 0.1000 | 10000 | 1994 | 5.02 | 1716 | 8606 |
| AXB | 0.1100 | 11000 | 624 | 17.63 | 746 | 13151 |
| SBI | 0.1100 | 11000 | 279 | 39.43 | 420 | 16559 |
| BJF | 0.1200 | 12000 | 5280 | 2.27 | 5967 | 13561 |
| BFS | 0.1200 | 12000 | 8870 | 1.35 | 11816 | 15986 |
| HLI | 0.0600 | 6000 | 678 | 8.85 | 686 | 6071 |
| SLI | 0.0500 | 5000 | 895 | 5.59 | 1007 | 5626 |
| Total | | 100000 | | | | 114714 |
| Actual Return: 14.71 % | | | | | | |

TABLE III. PREDICTED RETURN BY THE LSTM MODEL (FIN SERV SEC)

| Stock | Date: Jul 1, 2021 | | |
|---|---|---|---|
| | Pred Price | No of Stocks | Pred Value |
| HDB | 1484 | 16.12 | 23922 |
| HDF | 2462 | 0.58 | 1428 |
| ICB | 622 | 17.31 | 10767 |
| KTB | 1676 | 2.09 | 3503 |
| AXB | 742 | 10.83 | 8036 |
| SBI | 404 | 3.08 | 1244 |
| BJF | 5926 | 7.59 | 44978 |
| BFS | 11836 | 0.17 | 2012 |
| HLI | 670 | 1.11 | 744 |
| SLI | 987 | 13.71 | 13532 |
| Total | | | 110166 |
| Predicted Return: 10.17 % | | | |

Tables I – III depict the returns of two portfolio design approaches - optimum risk, and eigen, and the LSTM model predicted return for an investor who invested following the recommendations of the optimum risk portfolio. Fig. 2 shows the efficient frontier, the minimum risk portfolio, and the optimum risk portfolio of the financial services sector. As an illustration, Fig 3 depicts the plot of actual prices vs. corresponding predicted prices of the most significant stock in this sector, HDFC Bank, from Jan 1, 2021, to Jul 1, 2021.

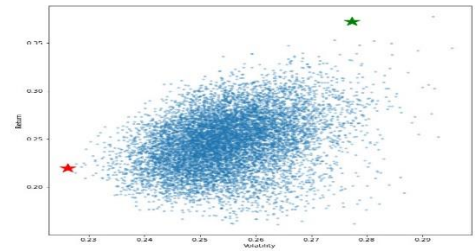

Fig. 2. The red and the green stars depict the min. risk portfolio and the opt. risk portfolio, respectively, for the financial services sector built on Jan 1, 2021. The *x*- and the *y*- axis plot the risk, return, respectively.

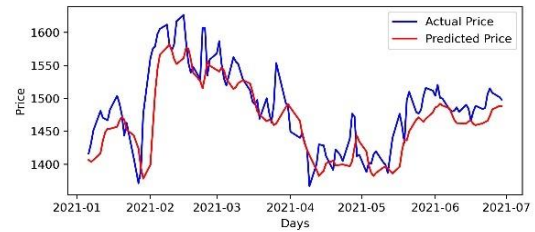

Fig. 3. Actual vs. predicted values of the HDFC Bank (HDB) stock as predicted by the LSTM model (Period: Jan 1, 2021, to Jul 1, 2021)

TABLE IV. ACTUAL RETURN OF OPT RISK PORTFOLIO (OIL & GAS SEC)

| Stock | Wts | Date: Jan 1, 2021 | | | Date: Jul 1, 2021 | |
|---|---|---|---|---|---|---|
| | | Amnt Invstd | Act Price | No of Stocks | Act Price | Act Value |
| RLI | 0.2338 | 23380 | 1988 | 11.76 | 2098 | 24674 |
| BPC | 0.0443 | 4430 | 382 | 11.60 | 463 | 5369 |
| ONG | 0.0479 | 4790 | 93 | 51.51 | 119 | 6129 |
| ATG | 0.1771 | 17710 | 377 | 46.98 | 969 | 45520 |
| IOC | 0.0084 | 840 | 92 | 9.13 | 108 | 986 |
| GAI | 0.0272 | 2720 | 124 | 21.94 | 153 | 3356 |
| IPG | 0.1186 | 11860 | 507 | 23.39 | 571 | 13357 |
| HPC | 0.0441 | 4410 | 221 | 19.95 | 296 | 5907 |
| PNL | 0.0370 | 3700 | 250 | 14.80 | 223 | 3300 |
| GJG | 0.2615 | 26160 | 378 | 69.18 | 675 | 46697 |
| Total | | 100000 | | | | 155295 |
| Actual Return: 55.30 % | | | | | | |

TABLE V. ACTUAL RETURN OF EIGEN PORTFOLIO (OIL & GAS SEC)

| Stock | Wts | Date: Jan 1, 2021 | | | Date: Jul 1, 2021 | |
|---|---|---|---|---|---|---|
| | | Amnt Invstd | Act Price | No of Stocks | Act Price | Act Value |
| RLI | 0.08 | 8000 | 1988 | 4.02 | 2098 | 8443 |
| BPC | 0.14 | 14000 | 382 | 36.65 | 463 | 16969 |
| ONG | 0.12 | 12000 | 93 | 129.03 | 119 | 15355 |
| ATG | 0.07 | 7000 | 377 | 18.57 | 969 | 17992 |
| IOC | 0.14 | 14000 | 92 | 152.17 | 108 | 16435 |
| GAI | 0.11 | 11000 | 124 | 88.71 | 153 | 13573 |
| IPG | 0.07 | 7000 | 507 | 13.81 | 571 | 7884 |
| HPC | 0.14 | 14000 | 221 | 63.35 | 296 | 18751 |
| PNL | 0.08 | 8000 | 250 | 32 | 223 | 7136 |
| GJG | 0.05 | 5000 | 378 | 13.23 | 675 | 8930 |
| Total | | 100000 | | | | 131468 |
| Actual Return: 31.47 % | | | | | | |

*B. Oil & Gas Sector*

The top ten stocks of the oil & gas sector and their weights (in percent) are Reliance Industries (RIL): 35.58, Bharat Petroleum Corporation (BPC), Oi & Natural Gas Corporation (ONG): 10.78, Adani Total Gas (ATG): 7.04, Indian Oil Corporation (IOC): 6.88, GAIL (GAI): 6.67, Indraprastha Gas (IPG): 4.90, Hindustan Petroleum Corporation (HPC): 4.70, Petronet LNG (PNL): 4.25, and Gujarat Gas (GJG): 2.85 [2]. Tables IV-VI show the returns of the two portfolios, optimum risk, the eigen, and the return predicted by the LSTM model.

Fig. 4 shows the efficient frontier, while Fig. 5 displays the actual and predicted prices of Reliance Ind (RLI), which is the leading stock of the oil & gas sector.

TABLE VI. Predicted Return by the LSTM Model (Oil & Gas Sec)

| Stock | Date: July 1, 2021 | | |
|---|---|---|---|
| | Pred Price | No of Stocks | Pred Value |
| RLI | 2069 | 11.76 | 24331 |
| BPC | 471 | 11.6 | 5464 |
| ONG | 119 | 51.51 | 6130 |
| ATG | 1065 | 46.98 | 50034 |
| IOC | 109 | 9.13 | 995 |
| GAI | 151 | 21.94 | 3313 |
| IPG | 535 | 23.39 | 12514 |
| HPC | 297 | 19.95 | 5925 |
| PNL | 224 | 14.8 | 3315 |
| GJG | 645 | 69.18 | 44621 |
| Total | | | 156642 |
| **Predicted Return: 56.64 %** | | | |

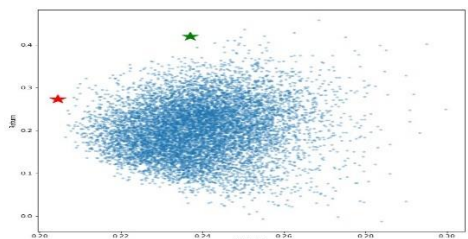

Fig. 4. The min. risk portfolio (the red star) and the opt. risk portfolio (the green star) for the oil & gas sector built on Jan 1, 2021 (The *x*-axis plots the risk, and the *y*-axis depicts the return)

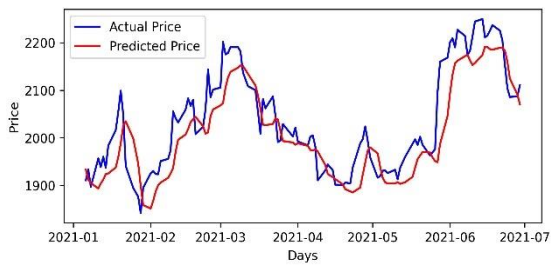

Fig. 5. Actual vs. predicted values of Reliance Industries (RLI) stock as predicted by the LSTM model (Period: Jan 1, 2021, to Jul 1, 2021)

TABLE VII. Actual Return of Opt Risk Portfolio (Pharma Sec)

| Stock | Wts | Date: Jan 1, 2021 | | | Date: Jul 1, 2021 | |
|---|---|---|---|---|---|---|
| | | Amnt Invstd | Act Price | No of Stocks | Act Price | Act Value |
| SPI | 0.0093 | 930 | 596 | 1.56 | 684 | 1067 |
| DRL | 0.0495 | 4950 | 5241 | 0.94 | 5559 | 5250 |
| DVL | 0.2506 | 25070 | 3849 | 6.51 | 4436 | 28893 |
| CPL | 0.0790 | 7900 | 827 | 9.55 | 978 | 9342 |
| LPN | 0.0316 | 3160 | 1001 | 3.16 | 1146 | 3618 |
| APH | 0.0068 | 680 | 928 | 0.73 | 968 | 709 |
| BCN | 0.2243 | 22430 | 466 | 48.13 | 406 | 19542 |
| CDH | 0.0770 | 7700 | 478 | 16.11 | 639 | 10294 |
| TPH | 0.1034 | 10340 | 2795 | 3.70 | 2924 | 10817 |
| AKL | 0.1684 | 16840 | 2951 | 5.71 | 3220 | 18376 |
| Total | | 100000 | | | | 107908 |
| **Actual Return: 7.91 %** | | | | | | |

## C. Pharmaceuticals Sector

The most significant stocks and their weights in this sector are Sun Pharmaceuticals Industries (SPI): 20.12, Dr. Reddy's Labs (DRL): 18.17, Divi's Labs (DVL): 15.50, Cipla (CPL): 13.62, Lupin (LPN): 7.63, Aurobindo Pharma (APH): 7.49, Biocon (BCN): 5.09, Cadila Healthcare (CDH): 4.56, Torrent Pharmaceuticals (TPH): 3.93, and Alkem Laboratories (AKL): 3.90 [2]. Tables VII-IX depict the returns produced by the two portfolios, optimum risk and eigen, and the LSTM-predicted return. Fig. 6 plots of the actual vs. predicted prices of the leading stock of the sector, Sun Pharmaceuticals (SPI).

TABLE VIII Actual Return of Eigen Portfolio (Pharma Sec)

| Stock | Wts | Date: Jan 1, 2021 | | | Date: Jul 1, 2021 | |
|---|---|---|---|---|---|---|
| | | Amnt Invstd | Act Price | No of Stocks | Act Price | Act Value |
| SPI | 0.12 | 12000 | 596 | 20.13 | 684 | 13772 |
| DRL | 0.11 | 11000 | 5241 | 2.10 | 5559 | 11667 |
| DVL | 0.10 | 10000 | 3849 | 2.60 | 4436 | 11525 |
| CPL | 0.11 | 11000 | 827 | 13.30 | 978 | 13008 |
| LPN | 0.12 | 12000 | 1001 | 11.99 | 1146 | 13738 |
| APH | 0.12 | 12000 | 928 | 12.93 | 968 | 12517 |
| BCN | 0.10 | 10000 | 466 | 21.46 | 406 | 8712 |
| CDH | 0.11 | 11000 | 478 | 23.01 | 639 | 14706 |
| TPH | 0.08 | 8000 | 2795 | 2.86 | 2924 | 8369 |
| AKL | 0.03 | 3000 | 2951 | 1.02 | 3220 | 3274 |
| Total | | 100000 | | | | 111288 |
| **Actual Return: 11.29 %** | | | | | | |

TABLE IX. Predicted Return by the LSTM Model (Pharma Sec)

| Stock | Date: Jul 1, 2021 | | |
|---|---|---|---|
| | Pred Price | No of Stocks | Pred Value |
| SPI | 675 | 1.56 | 1053 |
| DRL | 5429 | 0.94 | 5103 |
| DVL | 4281 | 6.51 | 27869 |
| CPL | 956 | 9.55 | 9130 |
| LPN | 1160 | 3.16 | 3666 |
| APH | 958 | 0.73 | 699 |
| BCN | 397 | 48.13 | 19108 |
| CDH | 632 | 16.11 | 10182 |
| TPH | 2928 | 3.70 | 10834 |
| AKL | 3132 | 5.71 | 17884 |
| Total | | | 105528 |
| **Predicted Return: 5.53 %** | | | |

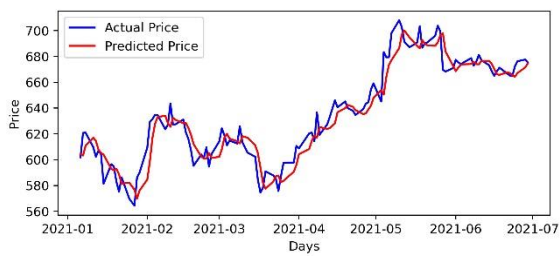

Fig. 6. Actual vs. predicted values of the Sun Pharmaceuticals (SIP) stock as predicted by the LSTM model (Period: Jan 1, 2021, to Jul 1, 2021)

TABLE X. Actual Return of Opt Risk Portfolio (PSU Bank Sec)

| Stock | Wts | Date: Jan 1, 2021 | | | Date: Jul 1, 2021 | |
|---|---|---|---|---|---|---|
| | | Amnt Invstd | Act Price | No of Stocks | Act Price | Act Value |
| SBI | 0.1981 | 19810 | 279 | 71.00 | 420 | 29822 |
| BOB | 0.0180 | 1800 | 65 | 27.69 | 86 | 2382 |
| PNB | 0.0561 | 5610 | 35 | 160.29 | 42 | 6732 |
| CNB | 0.0942 | 9420 | 133 | 70.83 | 154 | 10907 |
| UBI | 0.0084 | 840 | 32 | 26.25 | 39 | 1024 |
| BOI | 0.2322 | 23220 | 50 | 464.40 | 78 | 36223 |
| INB | 0.2320 | 23200 | 88 | 263.64 | 142 | 37436 |
| IOB | 0.0403 | 4030 | 11 | 366.36 | 27 | 9892 |
| CBI | 0.0025 | 240 | 14 | 17.86 | 28 | 500 |
| BMH | 0.1183 | 11830 | 14 | 845 | 25 | 21125 |
| Total | | 100000 | | | | 156043 |
| **Actual Return: 56.04 %** | | | | | | |

TABLE XI. Actual Return of Eigen Portfolio (PSU Bank Sec)

| Stock | Wts | Date: Jan 1, 2021 | Date: Jul 1, 2021 |
|---|---|---|---|

| | | Amnt Invstd | Act Price | No of Stocks | Act Price | Act Value |
|---|---|---|---|---|---|---|
| SBI | 0.11 | 11000 | 279 | 34.43 | 420 | 16559 |
| BOB | 0.11 | 11000 | 65 | 169.23 | 86 | 14554 |
| PNB | 0.11 | 11000 | 35 | 314.29 | 42 | 13200 |
| CNB | 0.11 | 11000 | 133 | 82.71 | 154 | 12737 |
| UBI | 0.11 | 11000 | 32 | 343.75 | 39 | 13406 |
| BOI | 0.11 | 11000 | 50 | 220.00 | 78 | 17160 |
| INB | 0.09 | 9000 | 88 | 102.27 | 142 | 14523 |
| IOB | 0.09 | 9000 | 11 | 818.18 | 27 | 22091 |
| CBI | 0.07 | 7000 | 14 | 500 | 28 | 14000 |
| BMH | 0.09 | 9000 | 14 | 643.86 | 25 | 16071 |
| **Total** | | **100000** | | | | **154301** |
| **Actual Return: 54.30 %** | | | | | | |

TABLE XII. PREDICTED RETURN BY THE LSTM MODEL (PSU BANK SEC)

| Stock | Date: Jul 1, 2021 | | |
|---|---|---|---|
| | Predicted Price | No of Stocks | Predicted Value |
| SBI | 414 | 34.43 | 14254 |
| BOB | 87 | 169.23 | 14723 |
| PNB | 43 | 314.29 | 13514 |
| CNB | 153 | 82.71 | 12655 |
| UBI | 38 | 343.75 | 13063 |
| BOI | 80 | 220.00 | 17600 |
| INB | 147 | 102.27 | 15034 |
| IOB | 26 | 818.18 | 21273 |
| CBI | 27 | 500 | 13500 |
| BMH | 25 | 643.86 | 16097 |
| **Total** | | | **151713** |
| **Predicted Return: 51.71 %** | | | |

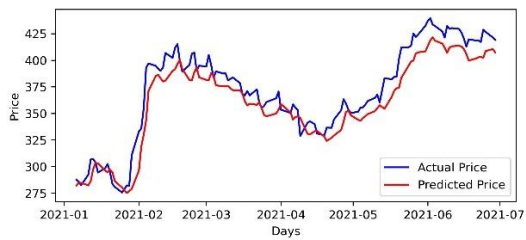

Fig. 7. Actual vs. the predicted values of State Bank of India (SBI) stock as predicted by the LSTM model (Period: Jan 1, 2021, to Jul 1, 2021)

*D. Public Sector Unit (PSU) Banks*

The top ten stocks and weights for the PSU Banks are State Bank of India (SBI): 28.03, Bank of Baroda (BOB): 18.80, Punjab National Bank (PNB): 14.78, Canara Bank (CNB): 13.70, Union Bank of India (UBI): 4.66, Bank of India (BOI): 4.51, Indian Bank (INB): 3.47, Indian Overseas Bank (IOB): 3.26, Central Bank of India (CBI): 3.02, and Bank of Maharashtra (BMH): 2.06 [2]. Tables X-XII exhibit the results of the portfolios and the LSTM model. Fig. 8 depicts the actual and predicted prices of the State Bank of India (SBI).

TABLE XIII. THE SUMMARY OF THE RESULTS

| Portfolio | Opt. Port Return (%) | Eigen Port Return (%) | LSTM Pred Return (%) |
|---|---|---|---|
| Fin Services | 11.15 | 14.71 | 10.17 |
| Oil & Gas | 55.30 | 31.47 | 56.64 |
| Pharma | 7.91 | 11.29 | 5.53 |
| PSU Banks | 56.04 | 54.30 | 51.71 |

Table XIII presents a summary of the results. Three important observations are as follows. First, except for the oil & gas sector, the returns yielded by the two portfolio strategies are quite similar. For the oil & gas sector, the opt risk portfolio has produced a significantly higher return than that its eigen counterpart. This is due to two stocks- ATG) and GJG. The optimum risk portfolio allocated larger capital to these stocks than the eigen portfolio and realized a substantially higher return. Second, the LSTM models' accuracy in prediction is found to be very high as it is observed that for all the four sectors, the returns forecasted by the LSTM model are very close to those actually produced by the optimum risk portfolios. Finally, the return of the PSU banks is found to be the highest, while the pharma sector yielded the lowest return.

VI. CONCLUSION

This paper has presented eight portfolios for four sectors of the Indian economy, taking into account the ten significant stocks from those sectors. For each sector, an optimum risk and an eigen portfolio are designed based on the historical prices of the stocks from Jan 1, 2016, to Dec 31, 2020. An LSTM model is also designed for predicting the stock prices with a forecast horizon of one day. After a hold-out period of six months, the actual and the predicted return of the portfolios are computed. The LSTM model is found to be highly accurate in predicting the future returns of the portfolios.

REFERENCES

[1] H. Markowitz, "Portfolio selection", *The Journal of Finance*, vol 7, no. 1, pp. 77-91, 1952.

[2] NSE Website: http://www1.nseindia.com

[3] S. Mehtab and J. Sen, "Stock price prediction using convolutional neural network on a multivariate time series", *Proc. of the 3rd NCMLAI'20*, Feb 2020, doi: 10.36227/techrxiv.15088734.v1.

[4] S. Mehtab and J. Sen, "A time series analysis-based stock price prediction using machine learning and deep learning models", *Int. J. of Business Forecasting and Mktg Int*, vol 6, no 4, pp. 272-335, 2021.

[5] J. Qiu and B. Wang, "Forecasting stock prices with long-short term memory neural network based on attention mechanism", *PLoS ONE*, vol. 15, no. 1, e0227222, 2020.

[6] J. Sen and S. Mehtab, "Accurate stock price forecasting using robust and optimized deep learning models", *Proc. of CONIT'21*, Jun 2021, India, doi: 10.1109/CONIT51480.2021.9498565.

[7] S. Metab and J. Sen, "A robust predictive model for stock price prediction using deep learning and natural language processing", *Proc. of 7th BAICONF*, Dec 2019, doi: 10.36227/techrxiv.15023361.v1.

[8] L. Shi, Z. Teng, L. Wang, Y. Zhang and A. Binder, "DeepClue: visual interpretation of text-based deep stock prediction", *IEEE Trans. On Knowledge and Data Engineering*, vol 31, no 31, pp. 1094-1108, 2019.

[9] N. Jing, Z. Wu, and H. Wang, "A hybrid model integrating deep learning with investor sentiment analysis for stock price prediction", *Expert Systems with Applications*, vol 178, 2019.

[10] J. Sen, S. Mehtab and A. Dutta, "Volatility modeling of stocks from selected sectors of the Indian economy using GARCH", *Proc of ASIANCON'21*, Aug 28-29, 2021, Pune, India, doi: 10.1109/ASIANCON51346.2021.9544977.

[11] S. Petchrompo, A. Wannakrairot, and A. K. Parlikad, "Puning pareto optimal solutions for multi-objective portfolios asset management", *European Journal of Operations Research*, May 2021. (In press).

[12] J. Sen and S. Mehtab, "Optimum risk portfolio and eigen portfolio: a comparative analysis using selected stocks from the Indian stock market, *Int. J. of Business Forecasting and Mktg. Intelligence*, Inderscience Journal, 2021 (In press).

[13] Y. Peng, P. H. M. Albuquerque, I-F, do Nascimento, and J. V. F. Machado, "Between nonlinearities, complexity, and noises: an application on portfolio selection using kernel principal component analysis", *Entropy*, vo;. 21, no. 4, p. 376, 2019.

[14] S. Almahdi and S. Y. Yang, "A constrained portfolio trading system using particle swarm algorithm and recurrent reinforcement learning", *Expert Systems with Applications*, vol. 130, pp. 145-156, 2019.

[15] Z. Wang, X. Zhang, Z. Zhang, and D. Sheng, "Credit portfolio optimization: a multi-objective genetic algorithm approach", *Bora Istanbul Review*, Jan 2021. (In press).

[16] A. Geron. *Hands-On Machine Learning with Scikit-Learn, Keras, and Tensorflow*, 2nd Edition, OReilly Media Inc, USA, 2019.